# CONSIDERATIONS FOR AN AC DIPOLE FOR THE LHC*

M. Bai, R. Calaga, W. Fischer, P. Oddo, BNL, Upton, Long Island, NY, 11973 U.S.A,
H. Schmickler, J. Serrano, CERN Laboratory, Geneva, Switzerland,
A. Jansson, M. Syphers, FNAL, Batavia, IL 60517, U.S.A.,
S. Kopp[#], R. Miyamoto, University of Texas, Austin, TX 78712, U.S.A.

*Abstract*

Following successful experience at the BNL AGS, FNAL Tevatron, and CERN SPS, an AC Dipole will be adopted at the LHC for rapid measurements of ring optics. This paper describes some of the parameters of the AC dipole for the LHC, scaling from performance of the FNAL and BNL devices.

## INTRODUCTION

The AC dipole [1,2] is a magnet whose current is oscillated sinusoidally in time so as to drive the beam transversely around a synchrotron, as shown in Figure 1. By tuning the magnet frequency to be near the tune times revolution frequency, the oscillations can be made to be coherent. The transverse oscillations can be used to perform measurements of the lattice of the machine, rapidly diagnose errors, and study non-linear optics. Such measurements are not unlike those made with a conventional pinger or kicker magnet, but the AC dipole can be operated for multiple turns, providing sustained coherent oscillations of the beam and improving the precision of the optics measurements. Further, the sinusoidal field can be ramped near-adiabatically in amplitude, resulting in no emittance growth of the beam [1]. Thus, measurements with an AC dipole are non-invasive, of significant advantage for the LHC because measurements can be made several times throughout the store or tune-up without having to dump the beam and refill the machine.

AC Dipole magnets have been used at the BNL RHIC [2], CERN SPS [3], and FNAL Tevatron [4]. Based upon the successful experiences of those systems, and based upon the similar relevant machine parameters for the Large Hadron Collider, we are developing a cost-effective AC Dipole system for commissioning of the LHC based upon commercial audio amplifiers and existing MKQ kicker magnets in the LHC ring.

## COMPARISON OF PARAMETERS

Several favourable parameters of the LHC, noted in Table 1, are amenable to the implementation of an AC Dipole which does not scale significantly from the systems employed at the Fermilab Tevatron and BNL RHIC. First, despite the fact that the LHC beam momentum is seven times that of the Tevatron, the typical machine beta functions are twice as large, making the beam transverse motion driven by the AC dipole correspondingly larger, and furthermore the beta function at the proposed LHC AC Dipole location is 258 m. Second, the required transverse motion does not scale with the beam momentum: in order to see a transverse displacement using a BPM system, we require only that the transverse displacement be greater than 4 times the beam transverse size, and the beam size is smaller at the LHC. Thus, the required magnetic field times the length of the LHC AC Dipole, $B_mL$, is 190 Gauss-meters, while that for the Tevatron and RHIC are only 140 and 142 Gauss-meters, respectively. Thus, if the magnet employed at FNAL as an AC Dipole [4,5] were utilized at the LHC (not foreseen), the required magnet current would be just 610 Amps, not significantly more than the 450 Amps at FNAL. Such considerations suggest that an AC Dipole at the LHC is practical, and does not impose technical challenges in proportion to the increase of beam momentum.

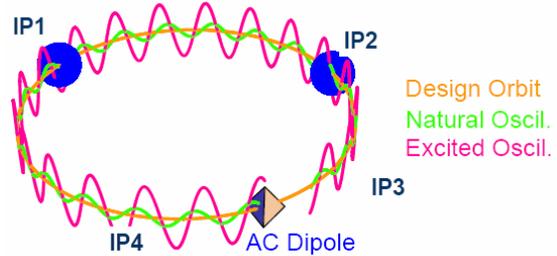

Figure 1: Schematic implementation of an AC dipole The dipole magnet provides a sinusoidally-varying kick over a period of many revolutions, while transverse motion of the beam around the ring is measured.

Table 1: Comparison of machine and AC Dipole parameters for the RHIC, Tevatron, and LHC. Shown are the revolution frequency, tune, AC Dipole magnet frequency, beam rigidity, maximum $\beta$ function around the ring (and also $\beta$ at the location of the AC Dipole), beam transverse size, required magnetic field strength to achieve $4\sigma$ oscillations of the beam, and required current in the magnet if a Tevatron-style pinger were used as the AC Dipole (see text).

| Machine (GeV) | RHIC (250) | Tevatron (980) | LHC (7000) |
|---|---|---|---|
| $f_{rev}$ [kHz] | 78 | 47.7 | 11 |
| $\nu$, $1-\nu$ | 0.71 | 0.58, 0.42 | 0.3, 0.7 |
| $f$ [kHz] | 55 | 20 | 3, 8, 14, 19 |
| $(B\rho)$ [Tm] | 833.9 | 3,300 | 23,000 |
| $\beta$ ($\beta_m$) [m] | 45 (11) | 80 (47) | 180 (258) |
| $\sigma$ [mm] | 0.75 | 0.5 | 0.3 |
| $B_mL$ [Gm] ($4\sigma$) | 142 | 140 | 190 |
| $I_{FNAL}$ [A] | 284 | 450 | 610 |

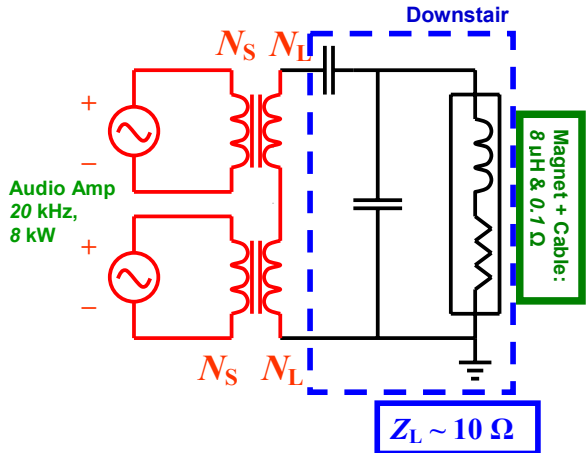

Figure 2: Schematic diagram of the AC Dipole scheme. A commercial audio amplifier(s) is coupled to the pinger magnet via a transformer. The low impedance of the magnet is matched to the optimal ~(8−10)Ω for the amplifier by means of a set of passive components in parallel with the magnet which form a resonant circuit. Multiple amplifiers may be joined to increase the power delivered to the magnet.

## MUSICAL DIPOLES

Two cost-saving design choices for the LHC AC Dipole system follow from the relatively low required driving frequencies. The revolution frequency is 11 kHz, and the operating tune is 0.3, so that the possible frequencies at which to drive the beam are relatively low. As noted in Table 1, several are even within the audio band: 3, 8, 14, and 19 kHz.

First, such low driving frequencies permit the use of a conventional pinger/kicker magnet as an AC Dipole, as was done for the Tevatron [4,5]. Though of higher inductance than the BNL air-core design which uses Litz-wires [6], the MKQ pinger-style magnet still has just 1.7 Ω inductive reactance when operated at 19 kHz. A further advantage of using an iron pinger magnet is its greater field-quality as compared to the air-core magnet. At the LHC, MKQ kickers will be used as AC Dipoles, with a relay switch enabling switching them from pinger to AC Dipole use during operations.

Second, the very low driving frequencies also permit use of low-cost commercial audio amplifiers to drive the AC Dipole, following the FNAL design [4,5]. The typical load for which such amplifiers deliver the maximum power is ~8 Ω, requiring their coupling to the magnet through a transformer and resonant circuit, as shown in Figure 2. The parallel capacitor maximizes the current through the magnet. Further, the low-$Q$ of the circuit arising from the magnet resistance is actually an advantage, because the amplitude of the current in the resonant circuit has a moderate fall-off around the nominal resonant frequency. We have studied several class D amplifiers such as the Crown T8000 (8 kW), the LabGrupen FP13000 (13kW), and are considering also the DIGAM K18 (18kW) models.

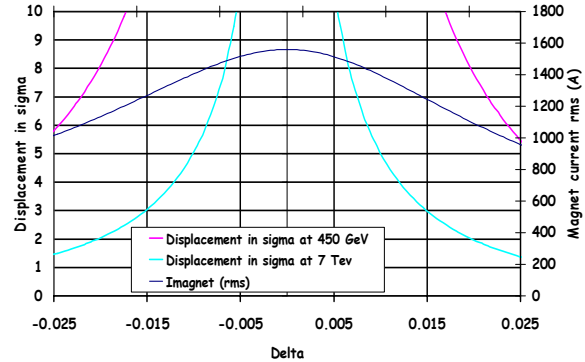

Figure 3: Simulation of the performance of the FP13000 amplifier driving the MKQ kicker as an AC dipole, operated at $f_m$ = 2.9 kHz. Shown is the current in the magnet (blue curve), beam displacement at 450 GeV/$c$ (pink curve), and beam displacement at 7 TeV/$c$ (cyan curve).

The amplifiers can be operated in stereo mode or in bridge mono mode, which doubles the power output in a single channel. Such motivates the transformer coupling and connecting the magnet to ground. It further permits increase in power by connection of additional amplifiers into the circuit, as shown in Figure 2. Though the current in the magnet grows only as the square root of the power, the increase is cost-effective. Such has been employed by BNL [6], and is under study with the Crown and Labgrupen amplifiers at FNAL.

Figures 3 and 4 show the simulated performance of the resonant circuit for the MQK kicker magnet connected to two LabGrupen FP13000 amplifier. Shown are the magnet current's amplitude, as well as the expected beam oscillation amplitude at 450 GeV and 7 TeV, as a function of $\delta=\nu-\nu_m$, where $\nu$ is the machine tune and $\nu_m$ is the magnet frequency tune. In this simulation, the magnets and cable inductance 1.8 µH and 2 mΩ were assumed, as well as a parallel capacitance of 760 µF and 120 µF at 2.9 and 8.2 kHz, respectively. The $Q$ values of the resonances are measured to be 6.8 and 10.2 at the two simulated driving frequencies. Typical operation points have been at $|\delta|=0.01$, at which point the beam displacement is still 4 times its transverse size when at 7 TeV energy.

## FREQUENCY TUNING

As noted above, the parallel capacitor in the resonant circuit is chosen to accomplish the current resonance at the desired driving frequency of the magnet. Thus, the fixed value of this capacitance fixes the resonant frequency; operation of the magnet at different frequencies forces us to operate off the resonance. Although the circuit $Q$ is relatively low, it may be seen that operation of the magnet at $\delta$~0.03 results in just ~1$\sigma$ oscillations of the beam at the machine flat top of 7 TeV.

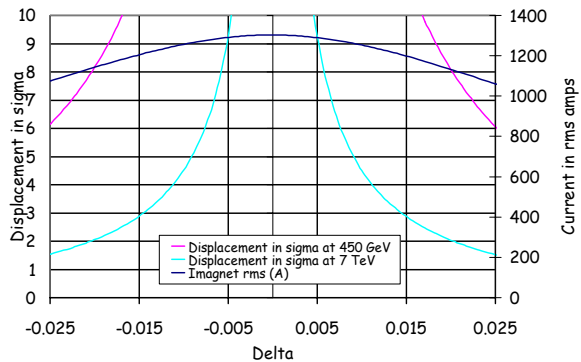

Figure 4: Simulation of the performance of the FP13000 amplifier driving the MKQ kicker as an AC dipole, operated at $f_m$=8.2kHz. Shown is the current in the magnet (blue curve), beam displacement at 450 GeV/$c$ (pink curve), and beam displacement at 7 TeV/$c$ (cyan curve).

As shown in [7,8], it is highly desirable to operate the AC Dipole at a variety of frequencies to properly infer machine lattice parameters from observed motion of the beam under the driving influence of the AC Dipole.

A wider tuning range with no change in the output current to the magnet can be achieved by means of a variable capacitor parallel to the magnet. We are exploring at BNL such a variable capacitor by means of connecting a switch in series with the parallel capacitor, as shown schematically in Figure 5. By opening and closing the switch with varying duty cycle, we vary the effective voltage across the capacitor, lowering the effective capacitance. The switching action can be viewed as an amplifier that does not deliver real power, and handles the reflected power.

Figure 6 shows the simulated magnetic field in a test magnet driven by the variable circuit of Figure 5. As can be seen, good uniformity in the field (and hence current) can be achieved for approximately ±200 Hz, corresponding to a change in $\delta$~0.02.

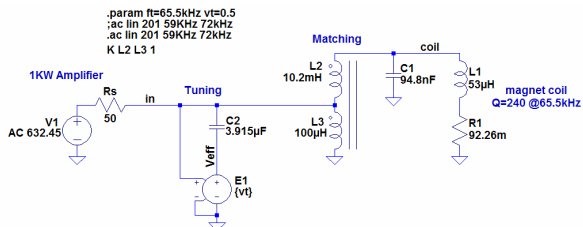

Figure 5: Schematic diagram being used to develop the variable parallel capacitor for the AC Dipole circuit. A MOSFET switch is connected across the capacitor in series. The switching action reduces the effective voltage across the capacitor, lowering the effective capacitance.

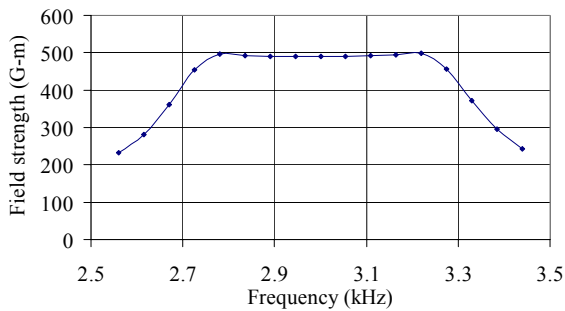

Figure 6: Simulation of the performance of a test AC Dipole magnet with equivalent circuit parameters using the variable capacitor scheme to vary the frequency around the nominal 2.9 kHz.

## SUMMARY

An AC Dipole system will be available for initial commissioning of the LHC at turn-on. Such devices have been extremely successful in performing machine diagnostics at BNL, SPS, and the Tevatron. Horizontal and Vertical AC Dipole magnets are foreseen for both rings. The initial system will follow the limited frequency range driving circuit like that employed at FNAL, and likely soon will be modified to use the variable switching capacitor design proposed by BNL.